\documentclass[]{article}
\usepackage{proceed2e}

\usepackage{times}

\usepackage{graphicx}
\usepackage{subfigure} 
\usepackage{amsmath,graphicx,mathtools}
\usepackage{amsfonts,amsthm}
\usepackage{amssymb}
\usepackage[dvipsnames]{xcolor}
\usepackage{dsfont}

\newtheorem{definition}{Definition}

\usepackage{natbib}
\usepackage{hyperref}

\begin{document} 
\title{Stochastic Portfolio Theory: \\ 
           A Machine Learning Perspective}

% The author names and affiliations should appear only in the accepted paper.
%
\author{ {\bf Yves-Laurent Kom Samo} \\
Machine Learning Research Group  \\
Oxford-Man Institute of Quantitative Finance \\
 University of Oxford \\
\textsc{ylks@robots.ox.ac.uk}\\
\And
{\bf Alexander Vervuurt}  \\
Mathematical Institute \\
Oxford-Man Institute of Quantitative Finance \\
 University of Oxford \\
\textsc{vervuurt@maths.ox.ac.uk}
}

\maketitle

\begin{abstract}
In this paper we propose a novel application of Gaussian processes (GPs) to financial asset allocation. Our approach is deeply rooted in \textit{Stochastic Portfolio Theory} (SPT), a stochastic analysis framework introduced by Robert Fernholz that aims at flexibly analysing the performance of certain investment strategies in stock markets relative to benchmark indices. In particular, SPT has exhibited some investment strategies based on company sizes that, under realistic assumptions, outperform benchmark indices with probability $1$ over certain time horizons. Galvanised by this result, we consider the inverse problem that consists of learning (from historical data) an optimal investment strategy based on any given set of trading characteristics, and using a user-specified optimality criterion that may go beyond outperforming a benchmark index. Although this inverse problem is of the utmost interest to investment management practitioners, it can hardly be tackled using the SPT framework. We show that our machine learning approach learns investment strategies that considerably outperform existing SPT strategies in the US stock market.
\end{abstract}

\section{INTRODUCTION} \label{sec:intro} 
%Y-L I have changed the intro (or abstract) only very little, I am happy with it as it is now but let's come back to it once all the other parts are done

Stochastic Portfolio Theory (SPT) is a relatively new stream in financial mathematics, initiated and largely developed by Robert~\cite{f02}. For surveys of the field, see \cite{fk09} and \cite{vervuurt}. Among many other things, SPT offers an alternative approach to portfolio selection, taking as its selection criterion to outperform the market index (for instance, the S\&P 500 index) with probability one. Investment strategies which achieve this are called \emph{relative arbitrages}, and have been constructed in certain classes of market models. 
The almost-sure comparison between the performance of certain portfolios and that of the market is facilitated by Fernholz's `master equation', a pathwise decomposition of this relative performance which is free from stochastic integrals. The foregoing master equation is the main strength of SPT portfolio selection, as it allows one to circumvent the challenges of explicit model postulation and calibration, as well as the (normative) no-arbitrage assumption, that are encountered in the classical approaches to portfolio optimisation. However, there remain several problems in and limitations to the SPT framework as it stands. 

First of all, the task of finding relative arbitrages under reasonable assumptions on the market model is difficult, since it is an inverse problem (this has also been noted by \cite{wong15}). Namely, given an investment strategy and market assumptions, one can check whether this strategy is a relative arbitrage (although this quickly becomes very hard for more complicated strategies), but the theory itself does not suggest such strategies. As such, the number of relative arbitrages that have been constructed explicitly remains very small. In a practical setting it would be preferable to invert the problem, and learn investment strategies from data using a user-specified performance criterion. In effect, most established investment managers will likely have a strong view on: i) what performance metric to use to evaluate their strategies, and ii) what values for the chosen metric they regard as being exceptional. The chosen performance metric may depart  from the excess return relative to the market index, for instance by adjusting for the risk taken. Similarly, outperforming the market index over a certain time horizon $[0,T]$ with probability $1$ might not be good enough for some practitioners, as investors might pull out following disappointing initial performances, leaving the investment manager unable to realise the long-term optimality. Whence, ideally one should aim at learning from market data what investment strategy is likely to perform exceptionally as per the user's view.

Secondly, several market imperfections are ignored in SPT; most notably, the possibility of bankruptcy is excluded. Since the constructed investment strategies typically invest heavily in small-capitalisation stocks, this poses a strong limitation on the real-world implementability of these portfolios. However, learning optimal investment strategies from the data copes well with bankruptcies as strategies investing in stocks that eventually fail will naturally be rejected as suboptimal. It also allows for the incorporation of transaction costs, which is theoretically challenging and has not yet been addressed in SPT.

Lastly, the SPT set-up has thus far been developed almost exclusively for investment strategies that are driven only by market capitalisations --- there have not yet been any constructions of relative arbitrages driven by other factors. Although this simplification eases theoretical analysis, it is a clear restriction as practitioners do consider many more market characteristics in order to exploit market inefficiencies.

We address all of these issues by adopting a Bayesian non-parametric approach. We consider a broad range of investment strategies driven by a function defined on an arbitrary space of trading characteristics (such as the market capitalisation), on which we place a Gaussian process (GP) prior. For a given strategy, the likelihood of it being `exceptional' is derived from a user-defined performance metric (e.g.~excess return to the market index, Sharpe ratio, etc) and values thereof that the practitioner considers `exceptional'. We then sample from the posterior of the GP driving the `exceptional' strategy using Monte Carlo Markov Chain (MCMC).

The rest of the paper is structured as follows. In section \ref{sec:spt} we provide a background on SPT. In section \ref{sec:ml} we present our model, and we illustrate that our approach learns strategies that outperform SPT alternatives in section \ref{sec:results}. Finally, we discuss our findings and make suggestions for future research in section \ref{sec:conclusion}.

%%%%%%%%%%%%%%%%%%%%%%%%%%%%%%%%%%%%%%%%%%%

\section{BACKGROUND} \label{sec:spt}
We give a brief introduction to SPT, defining the general class of market models within which its results hold, what the portfolio selection criterion is, and how strategies achieving this criterion are constructed.

%%%%%%%%%%%%%%%%
\subsection{THE MODEL} \label{sec:model}
%%%%%%%%%%%%%%%%
In SPT, the stock capitalisations are modelled as It\^o processes.\footnote{In the recent work \cite{kr16}, it has been shown that this can be weakened to a semimartingale model that even allows for defaults.} Namely, the dynamics of the $n$ positive stock capitalisation processes $X_i(\cdot),\,  i=1,\ldots,n\,$ are described by the following system of SDEs:
\begin{align} \label{model}
\mathrm{d}X_i(t) &= X_i(t)\bigg(b_i(t)\,\mathrm{d}t + \sum_{\nu=1}^d \sigma_{i\nu}(t)\, \mathrm{d}W_\nu(t)  \bigg),
\end{align}
for $t\geq0$ and $i=1,\ldots,n$. Here, $W_1(\cdot),\ldots,W_d(\cdot)$ are independent standard Brownian motions with $\, d \ge n$, and $X_i(0)>0,\ i=1,\ldots,n$ are the initial capitalisations. We assume all processes to be defined on a probability space $(\Omega,\mathcal{F},\mathbb{P})$, and adapted to a filtration $\mathbb{F} = \{ \mathcal{F} (t)\}_{0 \le t < \infty}\,$ that satisfies the usual conditions and contains the filtration generated by the  ``driving" Brownian motions. We refer the reader to \cite{ks88} for a reference on stochastic calculus.

The \emph{rates of return}  $b_i(\cdot),\ i=1,\ldots,n\,$ and \emph{volatilities}  $\sigma(\cdot)=(\sigma_{i\nu}(\cdot))_{1\leq i\leq n,1\leq \nu\leq d}\,,$ are some unspecified $\mathbb{F}$-progressively measurable processes and are assumed to satisfy the integrability condition
\begin{equation}
\sum_{i=1}^n \int_0^T \Big( |b_i(t)| + \sum_{\nu=1}^d(\sigma_{i\nu}(t))^2 \Big)\,\mathrm{d}t<\infty,\quad\mathbb{P}\text{-a.s.},
\end{equation}
for all$\,\,T\in(0,\infty),$ and the \emph{non-degeneracy} condition
\begin{equation} \label{ND} \tag{ND$^\varepsilon$}
\exists \, \varepsilon>0: \ \ \xi^T\sigma(t)\sigma^T(t)\xi\geq \varepsilon ||\xi||^2,
\end{equation}
for all $\xi\in\mathbb{R}^n$ and $t\geq0$,\, $\mathbb{P}$-almost surely.

%%%%%%%%%%%%%%%%
\subsection{RELATIVE ARBITRAGE} \label{sec:ra}
%%%%%%%%%%%%%%%%

In this context, one studies investments in the equity market described by \eqref{model} using \emph{portfolios}. These are $\mathbb{R}^n$-valued and $\mathbb{F}$-progressively measurable processes $\pi(\cdot) = \big( \pi_1 (\cdot), \cdots, \pi_n (\cdot) \big)^T$, where $\pi_i(t)$ stands for the proportion of   wealth invested in stock $i$ at time $t$. 

We restrict ourselves to \emph{long-only} portfolios. These invest solely in the stocks, namely, they take values in the closure $\overline{\Delta^n_+}$ of the set
\begin{align}
\Delta^n_+ &= \big\{ x\in\mathbb{R}^n :\  x_1+\ldots+x_n=1,\nonumber\\
&\qquad\qquad 0<x_i<1,\ i=1,\ldots,n \big\};
\end{align}
in particular, there is no money market. Assuming without loss of generality that the number of outstanding shares of each firm is 1, the corresponding wealth process $V^\pi(\cdot)$ of an investor implementing $\pi(\cdot)$ is seen to evolve as follows (we normalise the initial wealth to 1):
\begin{equation}
\frac{\mathrm{d}V^\pi(t)}{V^\pi(t)}\,=\,\sum_{i=1}^n \pi_i(t)\, \frac{\mathrm{d}X_i(t)}{X_i(t)}\,,\qquad V^\pi(0)=1.
\end{equation}

In SPT one measures performance, for the most part, with respect to the market index. This is the wealth process $V^\mu(\cdot)$ that results from a buy-and-hold portfolio,  given by the vector process $\mu(\cdot) = \big( \mu_1(\cdot), \cdots, \mu_n (\cdot) \big)^T$ of \emph{market weights}
\begin{equation} \label{defmu}
\mu_i(t)\,:=\,\frac{X_i(t)}{X_1(t)+\ldots+X_n(t)}.
\end{equation}

\begin{definition} \label{def:arbitrage}
Let $T>0$. A \emph{strong relative arbitrage} with respect to the market over the time-horizon $[0,T]$ is a portfolio $\pi(\cdot)$ such that
\begin{equation} \label{ra}
\mathbb{P}\big(V^\pi(T) > V^\mu(T)\big)=1.
\end{equation}
An equivalent way to express this notion, is to say that the portfolio $\pi(\cdot)$ \emph{strongly outperforms} $\mu(\cdot)$ over the time-horizon $[0,T]$. 
\qed
\end{definition} 

Contrast the SPT approach to portfolio selection with other methods such as mean-variance optimisation (originally introduced by \cite{markowitz}) and expected utility maximisation (see for instance \cite{rog13}), where the optimisation of a certain performance criterion determines the portfolio. In SPT, any portfolio that outperforms the market in the sense of \eqref{ra} is a relative arbitrage, and the amount by which it outperforms the market is theoretically irrelevant. 

In practice, one clearly desires this relative outperformance to be as large as possible.
Attempts at optimisation over the class of strategies that satisfy \eqref{ra} have been made by \cite{fernk10}, \cite{fk11}, \cite{ruf11}, \cite{ruf13a}, and \cite{wong15}.
However, these results are highly theoretical and very difficult to implement. Our data-driven approach circumvents these theoretical complications by optimising a user-defined criterion over the class of functionally-generated portfolios, which we introduce below.

%%%%%%%%%%%%%%%%%%%%%%%
\subsection{FUNCTIONALLY-GENERATED PORTFOLIOS} \label{sec:master}
%%%%%%%%%%%%%%%%%%%%%%%

A particular class of portfolios, called \emph{functionally-generated portfolios} (or FGPs for short), was introduced and studied by \cite{f95}. 

Consider a function $\mathbf{G}\in C^2(U,\mathbb{R}_+)$, where $U$ is an open neighbourhood of $\Delta^n_+$
and such that $x\mapsto x_i\mathrm{D}_i\log \mathbf{G}(x)$ is bounded on $\, \Delta^n_+ \,$ for $i=1,\ldots,n$.  % \footnote{We write $\mathrm{D}_i$ for the partial derivative with respect to the $i^\text{th}$ coordinate, and $\mathrm{D}^2_{ij}$ for the second partial derivative with respect to the $i^\text{th}$ and $j^\text{th}$ coordinates.}
Then $\mathbf{G}$ is said to be the \emph{generating function} of the \emph{functionally-generated portfolio} $\pi(\cdot)$, given, for $i=1,\ldots,n\,$, by
\begin{equation} \label{fgp}
\frac{\pi_i(t)}{\mu_i(t)}=\frac{\mathrm{D}_i\mathbf{G}(\mu(t))}{\mathbf{G}(\mu(t))} + 1 - \sum_{j=1}^n \mu_j(t)\frac{\mathrm{D}_j \mathbf{G}(\mu(t))}{\mathbf{G}(\mu(t))}\,.
\end{equation}
Here, we write $\mathrm{D}_i$ for the partial derivative with respect to the $i^\text{th}$ variable, and we will write $\mathrm{D}^2_{ij}$ for the second partial derivative with respect to the $i^\text{th}$ and $j^\text{th}$ variables. 
Theorem 3.1 of \cite{f95} asserts that the performance of the wealth process corresponding to $\pi(\cdot)$, when measured relative to the market, satisfies the $\mathbb{P}$-almost sure decomposition (often referred to as ``Fernholz's master equation")
\begin{equation} \label{master} 
\log\left(\frac{V^\pi(T)}{V^\mu(T)} \right) =\log \left(\frac{\mathbf{G}(\mu(T))}{\mathbf{G}(\mu(0))} \right) + \int_0^T \mathfrak{g}(t)\,\mathrm{d}t\,,
\end{equation}
where the quantity 
\begin{equation} \label{drift}
\mathfrak{g}(t) := -\sum_{i,j=1}^n \frac{\mathrm{D}_{ij}^2 \mathbf{G}(\mu(t))}{2\mathbf{G}(\mu(t))}\mu_i(t)\mu_j(t)\tau^\mu_{ij}(t)
\end{equation} 
is called the \emph{drift process} of the portfolio $\pi(\cdot)$. Here, we have written $\tau^\mu_{ij}(\cdot)$ for the \emph{relative covariances}; denoting by $e_i$ the $i$\textsuperscript{th} unit vector in $\mathbb{R}^n$, these are defined for $1\leq i,j\leq n$ as
\begin{equation} 
\tau^\mu_{ij}(t) := \big(\mu(t)-e_i\big)^T\sigma(t)\sigma^T(t)\big(\mu(t)-e_j\big).
\end{equation}

Under suitable conditions on the market model \eqref{model}, the left hand side of master equation \eqref{master} can be bounded away from zero for sufficiently large $T>0$, thus proving that $\pi(\cdot)$ is an arbitrage relative to the market over $[0,T]$. Several FGPs have been shown to outperform the market this way --- see \cite{f02}, \cite{fkk05}, \cite{fk05}, \cite{bf08}, \cite{fk09}, \cite{pic13}, and \cite{vk15}. In fact, \cite{palw14} prove that any relative arbitrage with respect to the market is necessarily of the form \eqref{fgp}, if one restricts $\pi(\cdot)$ to be a functional of the current market weights only.
% \cite{palw13}?

\cite{strong13} proves a generalisation of \eqref{master} for portfolios which are deterministic functions not only of the market capitalisations, but also of other observable quantities. Namely, let $\mathbf{x}(t)=(\mu(t),F)^T$, with $F$ a continuous, $\mathbb{R}^k$-valued, $\mathbb{F}$-progressively measurable process of finite variation, and let $\mathcal{H} \in C^{2,1}(\mathbb{R}^n\times\mathbb{R}^k, \mathbb{R}_+)$. By an application of Theorem 3.1 of \cite{strong13}, for any portfolio
\begin{equation} \label{fgpstrong}
\frac{\pi_i(t)}{\mu_i(t)}=\frac{\mathrm{D}_i\mathcal{H}(\mathbf{x}(t))}{\mathcal{H}(\mathbf{x}(t))} + 1 - \sum_{j=1}^n \mu_j(t)\frac{\mathrm{D}_j \mathcal{H}(\mathbf{x}(t))}{\mathcal{H}(\mathbf{x}(t))}\,,
\end{equation}
for $i=1,\ldots,n$, the following master equation holds:
\begin{align} \label{masterstrong} 
\log\left(\frac{V^\pi(T)}{V^\mu(T)} \right) &=\log \left(\frac{\mathcal{H}(\mathbf{x}(T))}{\mathcal{H}(\mathbf{x}(0))} \right) + \int_0^T \tilde{\mathfrak{g}}(t)\,\mathrm{d}t\, \\
&\quad -\int_0^T \sum_{l=1}^k \mathrm{D}_{n+l} \log \mathcal{H}(\mathbf{x}(t))\, \mathrm{d}F_l(t)\,.\nonumber
\end{align}
Here (compare with \eqref{master} and \eqref{drift})
\begin{equation}
\tilde{\mathfrak{g}}(t) := -\sum_{i,j=1}^n \frac{\mathrm{D}_{ij}^2 \mathcal{H}(\mathbf{x}(t))}{2\mathcal{H}(\mathbf{x}(t))}\mu_i(t)\mu_j(t)\tau^\mu_{ij}(t)\,.
\end{equation} 

Although explicit in its decomposition, the modified master equation \eqref{masterstrong} has so far not been applied in the literature. It is very difficult and unclear how to postulate in what way such `extended generating functions' $\mathcal{H}$ should depend on market information, and what additional covariates to use. It is thus of interest to develop a methodology that makes suggestions for what functions $\mathcal{H}$ to use, and extracts from market data which signals are significant. % We overcome this problem by adopting a Bayesian non-parametric approach --- see Section \ref{sec:ml}.

\subsection{DIVERSITY-WEIGHTED PORTFOLIOS} \label{sec:dwpp}
One of the most-studied FGPs is the \emph{diversity-weighted portfolio} (DWP) with parameter $p\in\mathbb{R}$, defined in (4.4) of \cite{fkk05} as
\begin{equation} \label{dwp}
\pi^{(p)}_i(t)\,:=\,\frac{(\mu_i(t))^p}{\, \sum_{j=1}^n (\mu_j(t))^p\,}\,,\quad i=1,\ldots,n.
\end{equation}
In Eq.~(4.5) of \cite{fkk05} it was shown that this portfolio is a relative arbitrage with respect to $\mu(\cdot)$ over $[0,T]$ for any $p\in(0,1)$ and $T>2\log n/(\varepsilon \delta p)$, under the condition \eqref{ND}, and that of \emph{diversity} \eqref{D}, introduced below. The latter says that no single company's capitalisation can take up more than a certain proportion of the entire market, which can be observed to hold in real markets;
\begin{equation} \label{D} \tag{D$^\delta$}
\exists \, \delta \in(0,1): \ \ \mathbb{P} \Big( \max_{\substack{1\leq i \leq n\\ t\in[0,T]}}\mu_i(t)<1-\delta \Big) \,=\,1\,.
\end{equation}
%Models of the form \eqref{model}, which satisfy the property \eqref{D} and other desirable properties, were explicitly shown to exist in Remark 6.2 of \cite{fkk05}. Other constructions have been proposed by \cite{or06} and \cite{sar14}.

In \cite{vk15}, this result was extended to the DWP with \emph{negative} parameter $p$, and several variations of this portfolio were shown to outperform the market over sufficiently long time horizons and under suitable market assumptions. A simulation using real market data supported the claim that these portfolios have the potential to outperform the market index, as well as their positive-parameter counterparts. Our results strongly confirm this finding, as well as computing the optimal parameter $p$ --- see section \ref{sec:results}.

%\subsection{GAMMA-WEIGHTED PORTFOLIOS} \label{sec:gwpp}

%Another portfolio that was put forward in \cite{vk15} is the so-called Gamma-weighted portfolio ($\Gamma$WP), namely the portfolio $\pi(\cdot)$ with weights
%\begin{equation}
%\pi_i(t)=\frac{\mu_i(t)^{a-1}e^{-b\mu_i(t)}}{\sum_{j=1}^n \mu_j(t)^{a-1}e^{-b\mu_j(t)}}\,, \qquad i=1, \ldots, n,
%\end{equation}
%and $a,b>0$ some constants. This portfolio attempts to avoid the problem of investing a large proportion of wealth in crashing stocks, which is encountered in the case of the diversity-weighted portfolio with negative parameter (since $\pi^{(p)}_i(t)\to1$ as $\mu_i(t)\to0$ for $p<0$), while maintaining the desired properties of that portfolio for mid- and large capitalisations.
% reiterate the 3 objectives

%%%%%%%%%%%%%%%%%%%%%%%%%%%%%%%%%%%%%%%%%%%

\section{SOLVING THE INVERSE PROBLEM} \label{sec:ml}
We consider solving the inverse problem of SPT: given some investment objective, how to go about learning a suitable trading strategy from the data? In doing so, we aim for a method that:
\begin{enumerate}
\item Learns from a large class of candidate investment strategies to uncover possibly intricate strategies from the data, typically by making use of non-parametric generative models for the generating functions;
\item Leverages additional sources of information beyond market capitalisations to uncover better investment strategies;
\item Works irrespective of the practitioner's investment objective (e.g.~achieving a high Sharpe Ratio, outperforming alternative benchmark indices, etc).
\end{enumerate}

\subsection{MODEL SPECIFICATION} %\label{sec:setup}
Let $\mathcal{X} \subset \mathbb{R}^d$ be a set of trading characteristics, for some $d\geq1$. We consider long-only portfolios of the form
\begin{equation} \label{pilogf}
\pi_i^f(t)=\frac{f\left(\mathbf{x}_i(t)\right)}{\sum_{j=1}^n f\left(\mathbf{x}_j(t)\right)}\,,\qquad i=1\ldots,n,
\end{equation}
for some continuous function $f:\mathcal{X}\to\mathbb{R}_+$. 

The idea behind this choice of investment portfolios is grounded in the fact that in practice, an investment manager will often have a predefined set of characteristics that he uses to compare stocks, for instance company size, balance sheet variables, credit ratings, sector, momentum, market vs.~book value,  return on assets, management team, online sentiment,  technical indicators, `beta', etc. The investment manager will typically choose trading characteristics so that they are informative enough to unveil market inefficiencies. Moreover, two stocks that have `similar' characteristics will receive `similar' weights. 

This approach includes as special cases all functionally-generated portfolios in the SPT framework, and in particular the diversity-, entropy- and equally-weighted, as well as market, portfolios. Our more general setting allows for any set of trading characteristics.

The trading opportunities in our framework are revealed through the time evolving trading characteristics $\mathbf{x}_i(t)$, and the investment map $f$ fully determines how to go about seizing these opportunities. Whence, learning an investment strategy in our framework is equivalent to learning an investment map $f$. To do so, we consider two families of functions. Firstly, galvanised by the theoretical results of SPT, we consider the case where $\mathcal{X} = \mathbb{R}_+$ is the set of market weights, and we take $f$ to be of the parametric form
\begin{equation} \label{dwpfct} %please do not change label, I reference to this eqn
f:\mu\mapsto \mu^p,
\end{equation}
for $p\in\mathbb{R}$, which corresponds to the diversity-weighted portfolio (DWP, see section \ref{sec:dwpp}). Secondly, in order to capture more intricate trading patterns, and to allow for a more general set of trading characteristics $\mathcal{X} \subset \mathbb{R}^d$, we also consider an alternative non-parametric approach in which we take $\log f$ to be a path of a mean-zero Gaussian process with covariance function $k$
\begin{equation}
\log f \sim \mathcal{GP}(0,k(\cdot,\cdot)).
\end{equation}
 
%
%
%Here, $\mathcal{X}$ is the set in which the \emph{vector of characteristics} $\mathbf{x}(\cdot)=(\mu(\cdot),\ldots)$ takes values --- any additional information on a firm that possibly changes in time can be included here. We denote by $\mathcal{H}$ the family of functions $\log f$. We consider two cases for $\mathcal{H}$: parametric and non-parametric families of functions. 
%
%The parametric form we consider is 
%\begin{equation} \label{dwpfct} %please do not change label, I reference to this eqn
%f:\mu_i\mapsto \mu_i^p,
%\end{equation}
%which corresponds to the diversity-weighted portfolio (DWP, see Section \ref{sec:dwpp}). We put as prior on $p$ a uniform distribution on $[-4, 4]$, and obtain a parametric functional prior $p(\log f)$.
%
%In the non-parametric case, we take $\mathcal{H}$ to be a reproducing kernel Hilbert space (RKHS) with continuous reproducing kernel $k(\cdot,\cdot)$. We put i.i.d.~standard normals as priors on the coordinates of $\log f$ in the orthogonal basis $\{\sqrt{\lambda_i} \phi_i(.)\}_{i \in \mathbb{N}}$ of $\mathcal{H}$, where $(\lambda_i, \phi_i(.))_{i \in \mathbb{N}}$ is the sequence of eigenvalues and eigenfunctions of the integral operator associated to $k(\cdot,\cdot)$ in Mercer's theorem. This implies that $$\log f \sim \mathcal{GP}(0,k(\cdot,\cdot)),$$ and 
%\begin{equation}
%\big(\log f\big(\mathbf{x}_1(1)\big),\ldots, \log f\big(\mathbf{x}_i(t)\big), \ldots, \log f\big(\mathbf{x}_n(T)\big)\big)
%\end{equation}
%is a multivariate Gaussian for every $T\in\mathbb{N}$.

To learn `good' investment maps, we need to introduce an optimality criterion that encodes the user's investment objective. To do so, we consider a performance functional $\mathcal{P}_\mathcal{D}$ that maps the logarithm of a candidate investment map to the historical performance $\mathcal{P}_\mathcal{D}(\log f )$ of the portfolio $\pi^f(\cdot)$ as in Eq. \eqref{pilogf} over some finite time horizon, given historical data $\mathcal{D}$. An example performance functional is the annualised Sharpe Ratio, defined as
\begin{equation}\label{eq:sr}
\text{SR}(\pi) = \sqrt{B}\,\frac{\hat{\mathbb{E}}\left(\{r(1), \dots, r(T)\}\right)}{\hat{\mathbb{S}}\left(\{r(1), \dots, r(T)\}\right)}\,,
\end{equation}
where $r(t) = \sum_{i=1}^n r_i(t)\pi_i^f(t)$ is the return of our portfolio between time $t-1$ and time $t$, $r_i(t)$ is the return of the $i$-th asset between time $t-1$ and time $t$, $B$ represents the number of units of time in a business year (e.g.~$252$ if the returns are daily), and $\hat{\mathbb{E}}$ (resp.~$\hat{\mathbb{S}}$) denote the sample mean (resp.~sample standard deviation). Another example of a performance functional is the excess return relative to a benchmark portfolio $\pi^*$
\begin{align} \label{ER}
\text{ER}\left(\pi^f \vert \pi^*\right) &= \prod_{t=1}^T \left(1 + \sum_{i=1}^n r_i(t)\pi_i^f(t)\right) \nonumber \\
&\quad - \prod_{t=1}^T \left(1 + \sum_{i=1}^n r_i(t)\pi_i^*(t)\right).
\end{align}
The nature of $\mathcal{P}_\mathcal{D}$ (Sharpe ratio, excess return, etc) depends on the portfolio manager; we impose no theoretical restriction.

In the parametric case (Eq. (\ref{dwpfct})), $\mathcal{P}_\mathcal{D}(\log f )$ is effectively a function of one single variable $p$, and we can easily learn the optimal $p$ using standard optimisation techniques.

In many cases, however, it might be preferable to reason under uncertainty and be Bayesian. To do so, we express the investment manager's view as to what is a good performance through a likelihood model $p (\mathcal{D} \big| \log f)$, which we may choose to be a probability distribution on $\mathcal{P}_\mathcal{D}(\log f )$
\begin{equation}
\mathcal{L}\left( \mathcal{P}_\mathcal{D}(\log f )\right) := p \left(\mathcal{D} \big| \log f\right).
\end{equation}
This is perhaps the most important step of the learning process. Indeed, the Bayesian methods we will develop in the next section aim at learning investment maps that provide an appropriate trade-off between how \emph{likely} the map is in light of training data, and how \emph{consistent} it is with prior beliefs. This will only lead to a profitable investment map if `likely' maps satisfy the manager's investment objective in-sample and vice-versa. If one chooses the likelihood model such that likely maps are strategies that lose money, then our learning machines will learn strategies that lose money! 

Fortunately, it is very straightforward to express that likely investment maps are the ones that match a desired investment objective. For instance, we may use as likelihood model that, given a candidate investment map $f$, the extent to which it is good, or equivalently the extent to which it is `likely' to be the function driving the strategy we are interested in learning, is the same as the extent to which the Sharpe Ratio it generates in-sample comes from a Gamma distribution with mean $2.0$ and standard deviation $0.5$. The positive support of the Gamma distribution renders functions leading to negative in-sample Sharpe ratios of no interest, while the concentration of the distribution over the Sharpe Ratio around $2.0$ reflects both our target performance and some tolerance around it. The choice of mean ($2.0$) and standard deviation ($0.5$) of the Gamma reflects the risk appetite of the investment manager, while the vanishing tails properly reflect the fact that too high a performance $\mathcal{P}_\mathcal{D}(\log f )$ would likely raise suspicions and too low a performance would not be good enough.

To complete our Bayesian model specification, in the parametric case we place on $p$ a uniform prior on $[-8, 8]$.

\subsection{INFERENCE} %\label{sec:setup}
Throughout the rest of this paper we will use as performance functional the total excess --- transaction cost adjusted --- return (as defined in \eqref{ER}) relative to the equally weighted portfolio (EWP), which has constant weights 
\begin{equation} \label{EWP}
\pi_i^{\text{EWP}}(t) = \frac{1}{n}\,,\quad i=1,\ldots,n\,,\ \forall\, t\geq0.
\end{equation}
over the whole training period
\begin{equation}
 \mathcal{P}_\mathcal{D}(\log f ) = \text{ER}( \pi^f \vert \text{EWP}).
\end{equation}
We assume a $10$bps transaction cost upon rebalancing (i.e.~we incur a cost of $0.1\%$ of the notional for each transaction). It is well known to algorithmic (execution) trading practitioners that a good rule of thumb is to expect to pay $10$bps when executing an order whose size is $10\%$ of the average daily traded volume (ADV) on liquid stocks. Whence, this assumption is reasonable so long as the wealth invested in each stock does not exceed $10 \%$ ADV. When needed, we use as likelihood model 
\begin{equation}
\mathcal{L}\left( \mathcal{P}_\mathcal{D}(\log f )\right) = \gamma \left(\mathcal{P}_\mathcal{D}(\log f ); a, b\right),
\end{equation}
where we denote $\gamma(.; a, b)$ the probability density function of the Gamma distribution with mean $a$ and standard deviation $b$. As previously discussed, $a$ and $b$ need not be learned as they reflect the investment manager's risk appetite. In the experiments of the following section, we use $a=7.0$ and $b=0.5$. In other words, we postulate that the ideal investment strategy should be such that, starting with a unit of wealth, the terminal wealth over the training period should be on average $7.0$ units of wealth higher than the terminal wealth achieved by the equally weighted portfolio over the same trading horizon --- this is purposely greedy.

\textbf{Frequentist parametric:} The first method of inference we consider consists of directly learning the optimal parameter of the DWP by maximising $\mathcal{P}_\mathcal{D}(\log f )$ for $p \in [-8, 8]$. As a comparison, the typical range of $p$ considered in the SPT literature is $[-1, 1]$. To avoid any issue with local maxima, we proceed with brute force maximisation on the uniform grid with mesh size $0.05$.\footnote{This took no longer than a couple of seconds in every experiment that we ran.}

\textbf{Bayesian parametric:} The second method of inference we consider consists of using the \emph{Metropolis-Hastings} algorithm (\cite{Hastings70}) to sample from the posterior distribution over the exponent $p$ in the DWP case,
\begin{equation}
\label{eq:post_dwp}
p(p | \mathcal{D}) \propto \mathcal{L}\left( \mathcal{P}_\mathcal{D}(p)\right) \times \mathds{1}\left(p \in [-8, 8]\right),
\end{equation}
where we have rewritten $\mathcal{L}\left( \mathcal{P}_\mathcal{D}(\log f )\right)$ as $\mathcal{L}\left( \mathcal{P}_\mathcal{D}(p)\right)$ to make the dependency in $p$ explicit. We sample a proposal update $p^*$ from a Gaussian centred at the current exponent $p$ and with standard deviation $0.5$. The acceptance probability is easily found to be 
\begin{equation}
r = \min\left(1, \frac{\mathcal{L}\left( \mathcal{P}_\mathcal{D}(p^*)\right)}{\mathcal{L}\left( \mathcal{P}_\mathcal{D}(p)\right)}\mathds{1}\left(p^* \in [-8, 8]\right) \right).
\end{equation}
We note in particular that so long as $p$ is initialised within $[-8, 8]$, the indicator function in Eq.~\eqref{eq:post_dwp} will not cause problems to the Markov chain. We typically run $10,000$ MH iterations and discard the first $5,000$ as `burn-in'. We use the posterior mean exponent learned on training data to trade in our testing horizon following the corresponding DWP
\begin{equation}
\hat{f}(\mu) = \mu^{\mathbb{E}\left(p \vert \mathcal{D}\right)}.
\end{equation}

\textbf{Bayesian non-parametric:} The third method of inference we consider is Bayesian and non-parametric. We place a Gaussian process prior on $\log f$
\begin{equation}
\log f \sim \mathcal{GP}(0,k(\cdot,\cdot)).
\end{equation}
Given the sizes of datasets we consider in our experiments (more than $3$ million training inputs --- $500$ assets over a $25$-year period), we approximate the latent function over a Cartesian grid. This approximation fits nicely with the quantised nature of financial data. We use as covariance function a separable product of Rational Quadratic (RQ) kernels  
\begin{equation}
k(x, y) = k_{0}^2 \prod_{i=1}^d  \left(1 + \frac{(x[i]-y[i])^2}{2 \alpha_i l_i^2} \right)^{-\alpha_i},
\end{equation}
where the hyper-parameters $k_0, l_i, \alpha_i >0$, on which we place independent log-normal priors are all to be inferred. We found the RQ kernel to be a better choice than the Gaussian kernel as it allows for `varying length scales'. Denoting by $\mathbf{f}$ the values of the investment map over the input grid, we prefer to work with the equivalent whitened representation
\begin{equation}
\log \mathbf{f} = L\mathbf{X},\quad  ~\mathbf{X} \sim \mathcal{N}(0, I),
\end{equation}
where $I$ is the identity matrix, $K = [k(\mathbf{x}_i, \mathbf{x}_j)]_{i,j \leq N}$ is the Gram matrix over all $N$ input points, $K = U D U^T$ is the Singular Value Decomposition (SVD) of $K$ and $L= UD^{\frac{1}{2}}$. We use a Blocked Gibbs sampler (\cite{Gibbs84}) to sample from the posterior 
\begin{align} \label{posterior}
&p\left(\log \mathbf{X},\log  k_{0},\{ \log l_i, \log \alpha_i\}_{i \leq d} \vert  \mathcal{D}\right) \propto \mathcal{L}\left( \mathcal{P}_\mathcal{D}(L\mathbf{X})\right)\nonumber \\ 
& \quad \times   p(\log \mathbf{X}) p(\log k_0) \prod_{i=1}^d p(\log l_i) p(\log \alpha_i)\,,
\end{align}
where we have rewritten $\mathcal{L}\left( \mathcal{P}_\mathcal{D}(\log f )\right)$ as $ \mathcal{L}\left( \mathcal{P}_\mathcal{D}(L\mathbf{X})\right)$ to emphasise that the likelihood is fully defined by $\mathbf{f} = L\mathbf{X}$. The whitened representation has two primary advantages. First, it is robust to ill conditioning as we may always compute $L$, even when $K$ is singular. Second, it creates a hard link between function values and hyper-parameters, so that updating the latter affects the likelihood $\mathcal{L}\left( \mathcal{P}_\mathcal{D}(\log f )\right)$, and therefore directly contributes towards improving the training performance $\mathcal{P}_\mathcal{D}(\log f )$: we found this to improve mixing of the Markov chain. Our Blocked Gibbs sampler alternates between updating $\log \mathbf{X}$ conditional on hyper-parameters, and updating the hyper-parameters (and consequently $L$) conditional on $\log \mathbf{X}$. For both steps we use the \emph{elliptical slice sampling} algorithm (\cite{murray2010}). The computational bottleneck of our sampler is the computation of the SVD of $K$, which we may do very efficiently by exploiting the separability of our kernel and the grid structure of the input space using standard Kronecker techniques (see for instance \cite{saatchi11}).

%%%%%%%%%%%%%%%%%%%%%%%%%%%%%%%%%%%%%%%%%%%

\section{EXPERIMENTS} \label{sec:results}
The universe of stocks we consider in our experiments are the constituents of the S\&P 500 index, accounting for changes in index constituents. We rebalance our portfolios on a daily basis. At the end of each trading day, we determine our target portfolio for the next day, which is acquired at the open of the next trading day. When the constituents of the index are due to change on day $t$, our target portfolio at the end of day $t-1$ relates to the constituents of the index on day $t$ (which would indeed be known to the market on day $t-1$). As previously discussed, we assume that each transaction incurs a charge of $0.1\%$ of its notional value. The returns we use account for corporate events such as dividends, defaults, M\&A's, etc. Our data sources are the CRSP and Compustat databases, and we use data from 1 January 1992 to 31 December 2014.

In our first experiment, we aim to illustrate that the approaches we propose in this paper \emph{consistently} and considerably outperform SPT alternatives over a wide range of market conditions. We consider learning optimal investment strategies as described in the previous section using 10 consecutive years worth of data and testing on the following 5 years. We begin on 1st January 1992 for the first training dataset, and roll both training and testing datasets by one year, which leads to a total of 9 pairs of training and testing subsets. We compare the equally-weighted portfolio (EWP), the market portfolio, the diversity-weighted portfolio where the exponent $p$ is learned by maximising the evaluation functional (DWP*), the diversity-weighted portfolio where the exponent $p$ is learned with MCMC (DWP), the Gaussian process approach using as trading characteristic the logarithm of the market weights (CAP), and the Gaussian process approach using as trading characteristics both the logarithm of the market weights and the return-on-assets (CAP+ROA). The return-on-assets (ROA) on day $t$ is defined as the ratio between the last net income reported by the company and last total assets reported by the company known on day $t$ --- we note that this quantity may not change on a daily basis but this does not affect our analysis. The rationale behind using the ROA as additional characteristics is to capture not only how big a company is, but also how well it performs relative to its size.

Table \ref{tab:unif} summarises the average over the 9 scenarios of the yearly in-sample and out-of-sample returns plus-minus two standard errors. It can be seen that all learned strategies do indeed outperform the benchmark (EWP) in-sample and out-of-sample. Moreover, the performance is greatly improved by considering non-parametric models, even when the only characteristic considered is the market weight. Analysing such families of strategies within the SPT framework would simply be mathematically intractable. Finally, it can be seen that adding more trading characteristics does indeed add value. Crucially, the CAP+ROA portfolio \emph{considerably} and \emph{consistently} outperforms the benchmark (EWP), both in-sample and out-of-sample.

\begin{figure} [th!]
\centering
\includegraphics[width=0.5\textwidth]{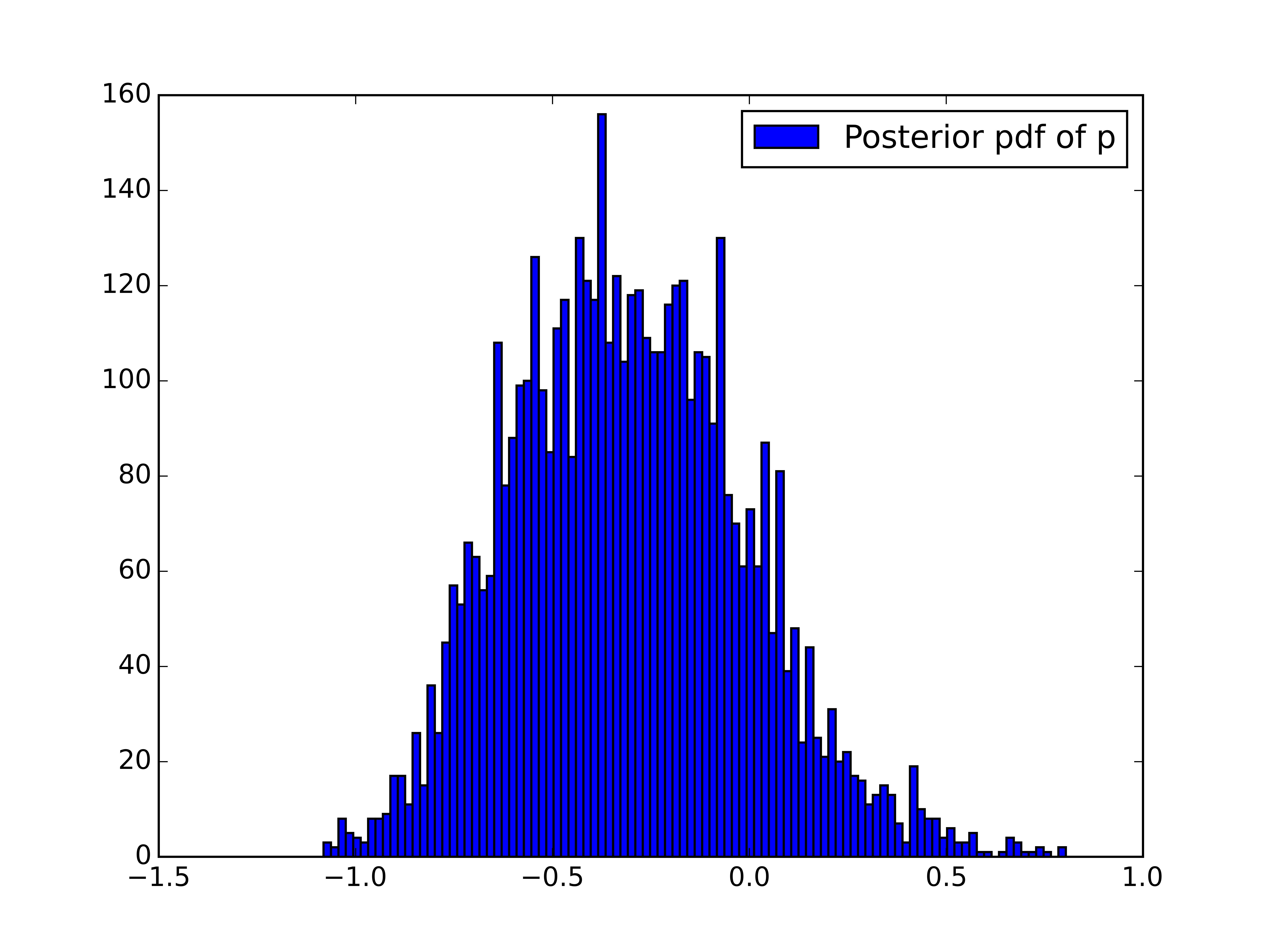}
\caption{Posterior distribution of the parameter $p$ of the diversity-weighted portfolio in our second experiment. The model was trained with market data between 1st January 1992 to 31st December 2005.} \label{fig:histo}
\end{figure}

\begin{figure} [h!]
\centering
\includegraphics[width=0.5\textwidth]{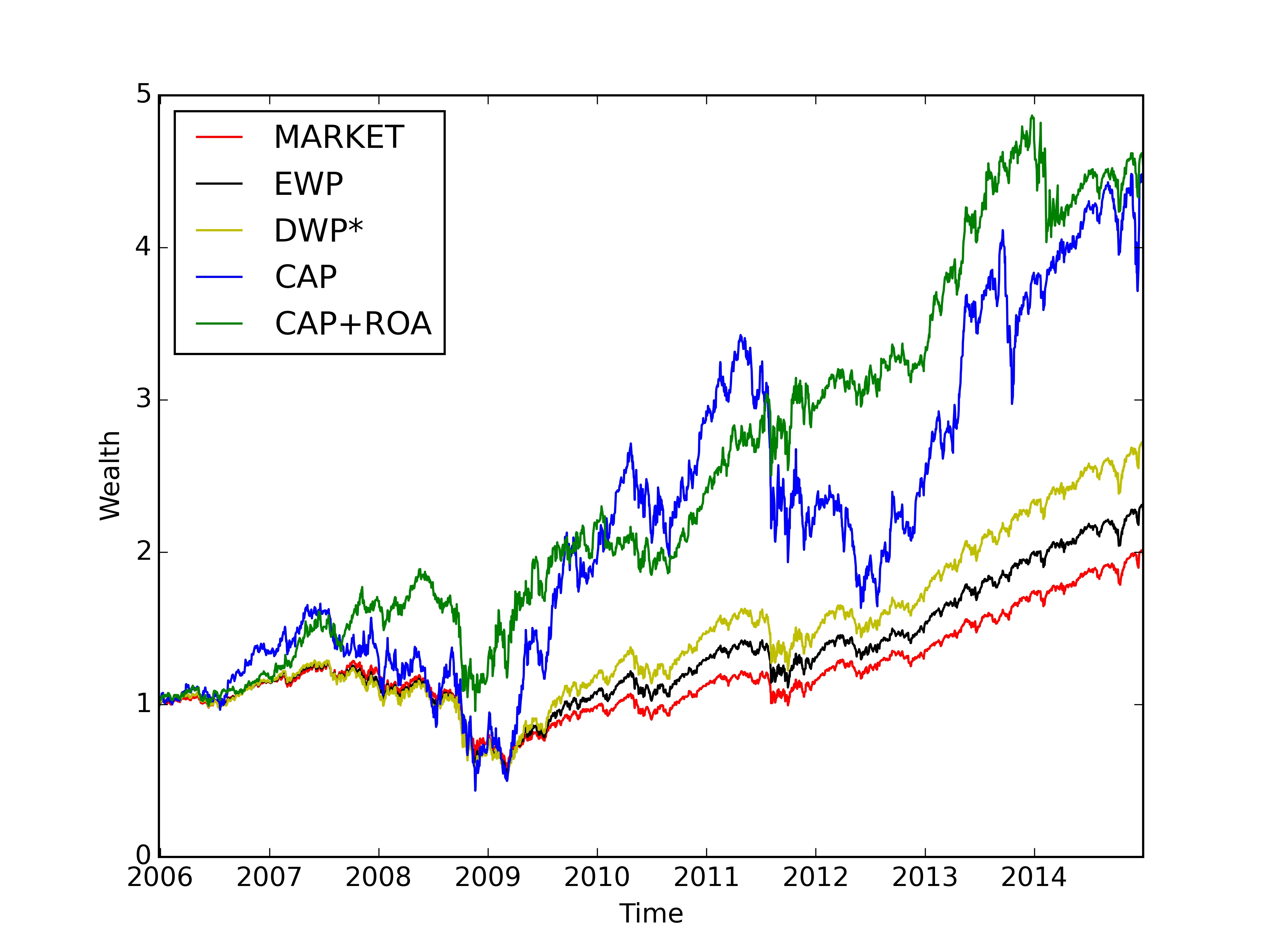}
\caption{Time series of out-of-sample wealth processes in our second experiment. Models were trained with market data between 1st January 1992 to 31st December 2005, and tested from 1st January 2006 to 31st December 2014.}
\label{fig:V}
\end{figure}

\begin{figure} [h!]
\centering
\includegraphics[width=0.4\textwidth]{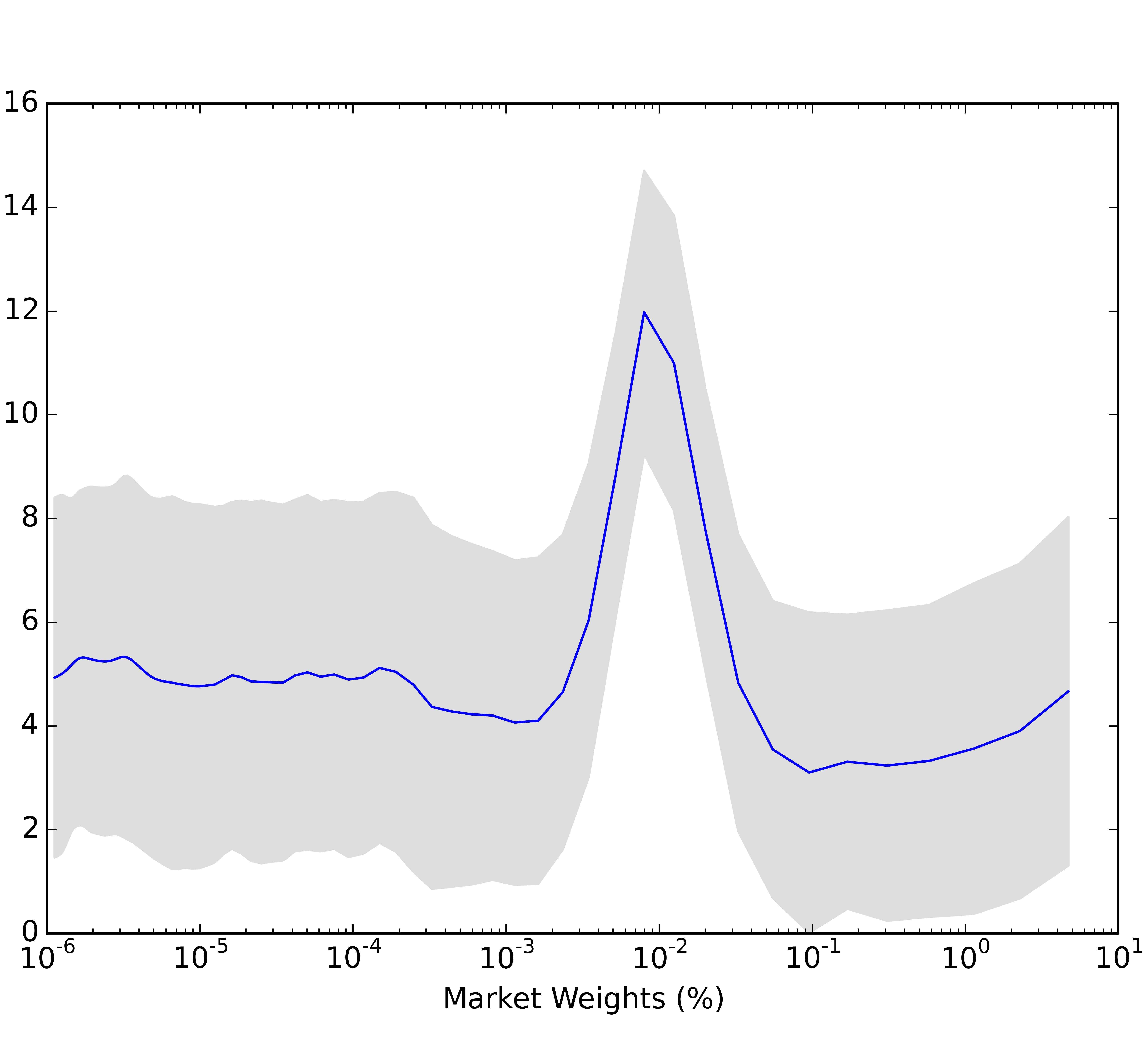}
\includegraphics[width=0.5\textwidth]{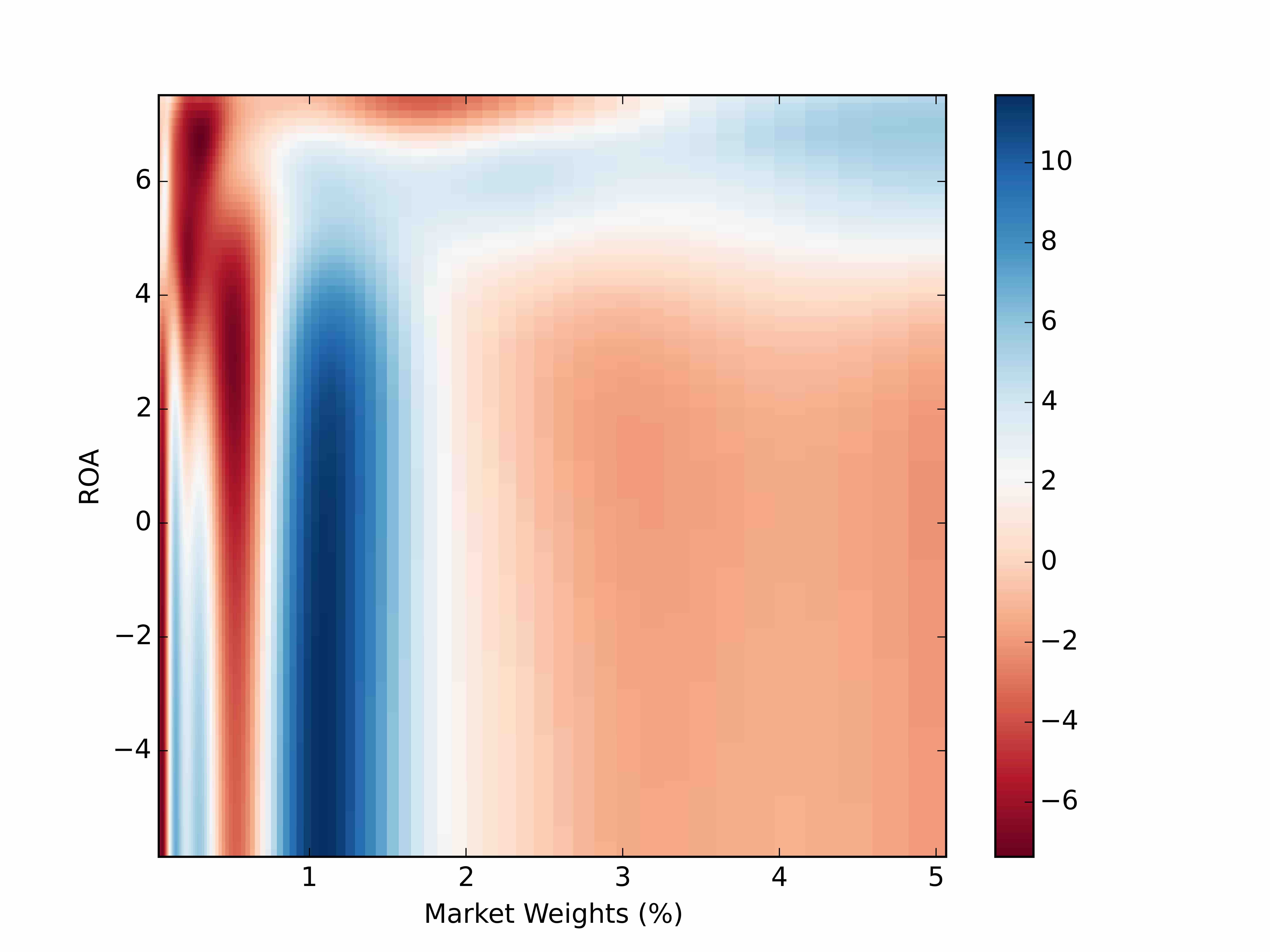}
\caption{Learned logarithm investment maps of the CAP portfolio (top) and the CAP+ROA portfolio (bottom) in our second experiment. In the case of the CAP portfolio, the credible band corresponds to $\pm$ $2$ posterior standard deviations.}
\label{fig:GPmap}
\end{figure}

\begin{table}[th]
\caption{Results of our first experiment on the consistency of our learning algorithms to varying market conditions. IS RET (resp.~OOS RET) are in-sample (resp.~out-of-sample) average (over the 9 runs in the experiment) yearly returns in \% $\pm$ two standard errors.} \label{tab:unif}
\begin{center}
\begin{tabular}{lcc}
\textsc{Portfolio} & \textsc{IS ret} (\%) & \textsc{OOS ret} (\%) \\
\hline \\
%\abovespace
\textsc{Market} & 8.56$\pm$1.62 & 6.23$\pm$2.07 \\
EWP & 10.56$\pm$1.67 & 8.99$\pm$1.85 \\
DWP* & 11.94$\pm$2.01 & 12.51$\pm$1.12\\
DWP & 11.91$\pm$1.99 & 12.50$\pm$1.11 \\
CAP & 26.54$\pm$2.38 & 22.05$\pm$2.89 \\
CAP+ROA & \textbf{56.18$\pm$7.35} & \textbf{25.14$\pm$2.58}\\
%GP \textsc{mom} & & 14.23$\pm$2.33 & \textbf{0.634$\pm$0.121} \\
\end{tabular}
\end{center}
\end{table}
In our second experiment, we aim to illustrate that our approaches are robust to financial crises. To do so, we train our model using data from 1 January 1992 to 31 December 2005, and test the learned strategy between 1 January 2006 and 31 December 2014, which includes the 2008 financial crisis. We compare the same investment strategies as before. The posterior distribution over the exponent $p$ in the Bayesian parametric method is illustrated in Figure \ref{fig:histo}. The learned posterior mean investment maps are illustrated in Figure \ref{fig:GPmap}. In Table \ref{tab:prelim} we provide in-sample and out-of-sample average yearly returns as well as out-of-sample Sharpe ratios. Once again, it can be seen that: i) all learned portfolios do indeed outperform the benchmark (EWP) in-sample and out-of-sample, ii) non-parametric methods outperform parametric methods, and iii) adding the ROA as an additional characteristic does indeed add value. These conclusions hold true not only in absolute terms (returns) but also after adjusting for risk (Sharpe Ratio). A more granular illustration of how our method performs during the 2008 financial crisis can be seen in the time series of total wealth provided in Figure \ref{fig:V}. It turns out that the ROA does not only improve the return out-of-sample, but it also has a `stabilising effect' in that the volatility of the wealth process is considerably reduced.

Finally, it is also worth stressing that the shape of the learned investment map in the two-dimensional case (Figure \ref{fig:GPmap}) suggests that the investment strategy uncovered by our Bayesian nonparametric approach can hardly be replicated with a parametric model. Once again, it would be near impossible to derive analytical results pertaining to such a portfolio within the SPT framework.

%%%%%%%%%%%%%%%%%%%

\begin{table}[h]
\caption{Results of our second experiment on the robustness of the proposed approaches to financial crises. Returns (RET) are yearly equivalents (in \%) of the total returns over the whole testing period. The annualised Sharpe Ratio (SR) is as per Eq.~\eqref{eq:sr}. IS (resp.~OOS) stands for in-sample (resp.~out-of-sample).} \label{tab:prelim}
\begin{center}
\begin{tabular}{lccc}
\textsc{Portfolio} & IS RET (\%) & OOS RET (\%) & OOS SR \\
\hline \\
%\abovespace
\textsc{Market} & 9.60 & 7.90 & 0.47\\
EWP & 13.46 & 9.60 & 0.51 \\
DWP* & 14.62 & 11.74 & 0.56 \\
DWP & 14.62 & 11.38 & 0.55 \\
CAP & 16.49 & 18.01 & 0.60\\
CAP+ROA & \textbf{37.54} & \textbf{18.33} & \textbf{0.72} \\
%GP \textsc{mom} & 15.54 & 11.24 & 0.493 \\
\end{tabular}
\end{center}
\end{table}

%wealth processes

%GP

\section{CONCLUSION \& DISCUSSION} \label{sec:conclusion}
The inverse problem of \emph{stochastic portfolio theory} (SPT) is the following problem: given a user-defined portfolio selection criterion, how does one go about constructing suitable investment strategies that meet the desired investment objective? This problem is extremely challenging to solve within the SPT framework. We propose the first solution to the inverse SPT problem and we demonstrate empirically that the proposed methods consistently and considerably outperform standard benchmarks, and are robust to financial crises. 

Unlike the SPT framework, our methods are based solely on historical data rather than stochastic calculus. This allows us to consider a very broad class of candidate investment strategies that includes all SPT strategies as special cases, but crucially contains many investment strategies that cannot be analysed in the SPT framework. Unlike the SPT framework, which almost exclusively considers outperforming the market portfolio using investment strategies that are solely based on market weights, our proposed approach can cope with virtually any user-defined investment objective and can exploit any arbitrary set of trading characteristics. We empirically demonstrate that this added flexibility allows us to uncover more subtle patterns in financial markets, which results in greatly improved performance.

Although the Gaussian process in our model was approximated to be piecewise constant on a grid, there is no theoretical or practical obstacle in using an alternative approximation such as sparse Gaussian processes (\cite{FTCI}) or string Gaussian processes (\cite{samo_sgp}). Our method may be extended to learn even subtler patterns using the non-stationary general purpose kernels of \cite{samo15gen}. Our work may also be extended to allow for long-short investment strategies (i.e.~strategies that allow short-selling). Finally, it would be interesting to develop an online extension of our work that would capture temporal changes in market dynamics. 

\subsubsection*{Acknowledgements}
Yves-Laurent would like to acknowledge support from the Oxford-Man Institute of Quantitative Finance. Alexander gratefully acknowledges PhD studentships from the Engineering and Physical Sciences Research Council, Nomura, and the Oxford-Man Institute of Quantitative Finance. Wharton Research Data Services (WRDS) was used in preparing the data for this paper. This service and the data available thereon constitute valuable intellectual property and trade secrets of WRDS and/or its third-party suppliers.

\bibliographystyle{plainnat}
\bibliography{sptrefs}

\end{document}